\documentclass[showpacs,preprintnumbers,amsmath,amssymb]{revtex4}
\usepackage{graphicx} 
\usepackage{dcolumn} 
\usepackage{bm} 
\usepackage{epsfig}
\usepackage{amsfonts}

\begin{document}

\title{A closer look at time averages of the logistic map at the edge of chaos}

\author{Ugur Tirnakli}
\email{ugur.tirnakli@ege.edu.tr}
\affiliation{Department of Physics, Faculty of Science, Ege University, 35100 Izmir, Turkey}

\author{Constantino Tsallis}
\affiliation{Centro Brasileiro de Pesquisas F\'\i sicas, R. Dr. Xavier Sigaud 150, 22290-180
Rio de Janeiro, RJ, Brazil\\
and\\
Santa Fe Institute, 1399 Hyde Park Road, Santa Fe, NM 87501, USA}

\author{Christian Beck}
\affiliation{School of Mathematical Sciences, Queen Mary, University of London, Mile End Road,
London E1 4NS, UK }

\date{\today}

\begin{abstract}
The probability distribution of sums of iterates of the logistic map at the edge of chaos has been recently
shown [see U. Tirnakli, C. Beck and C. Tsallis, Phys. Rev. E {\bf 75}, 040106(R) (2007)] to be numerically
consistent with a $q$-Gaussian, the distribution which, under appropriate constraints, maximizes the
nonadditive entropy $S_q$, the basis of  nonextensive statistical mechanics.
This analysis was based on a study of the tails of the distribution. We now check the entire distribution,
in particular its central part. This is important in view  of a recent $q$-generalization of the
Central Limit Theorem, which states that for certain classes of strongly correlated random variables
the rescaled sum approaches a $q$-Gaussian limit distribution. We numerically investigate for the
logistic map with a parameter in a small vicinity of the critical point under which conditions there is
convergence to a $q$-Gaussian both in the central region and in the tail region, and find a scaling law
involving the Feigenbaum constant $\delta$. Our results are consistent with a large number of already
available analytical and numerical evidences that the edge of chaos is well described in terms of the
entropy $S_q$ and its associated concepts.
\end{abstract}


\maketitle

One of the cornerstones of statistical mechanics and of probability theory  is the Central Limit Theorem (CLT).
It states that the sum of $N$ independent identically  distributed random variables, after appropriate
centering and rescaling, approaches a Gaussian distribution as $N\rightarrow\infty$. In general, this concept
lies at the very heart of the fact that many  stochastic processes in nature which consist of a sum of many
independent or nearly independent variables converge to a Gaussian process \cite{vKa,khinchin}. On the other hand,
there are also many other occasions in nature for which the limit distribution is not a Gaussian.
The common ingredient  for such systems is the existence of strong correlations between the random variables,
which prevent the limit distribution of the system to end up being a Gaussian. Recently, for certain classes of strong
correlations of this kind, it has been proved that the 
distribution of the rescaled sum approaches a $q$-Gaussian, which constitutes
a $q$-generalization of the standard CLT  \cite{tsa3,TsallisQueirosCATANIA,QueirosTsallisCATANIA,vignatplastinocentral2}.
This represents a progress  since the $q$-Gaussians are the distributions that optimize the nonadditive entropy
$S_q$ (defined to be $S_{q} \equiv \left(1- \sum_i p_i^q\right)/ \left(q-1\right)$), on which nonextensive
statistical mechanics is based \cite{Tsallis,TGM}. A $q$-generalized CLT was expected for several years
since the role of $q$-Gaussians in nonextensive statistical mechanics is pretty much the  same as that of
Gaussians in Boltzmann-Gibbs statistical mechanics. Therefore it is not  surprising at all to see $q$-Gaussians
replace the usual Gaussian distributions for those systems whose agents exhibit certain types of
strong correlations.

Immediately after these achievements, an increasing interest developed for checking these ideas and findings
in real and model systems whose dynamical properties make them  appropriate candidates to be analyzed along
these lines. Cortines and Riera have analysed  stock market index changes for a considerable range of time
delays using Brazilian financial  data \cite{cortines} and found that the histograms can be well approximated
by $q$-Gaussians with $q\approx 1.75$ (see Fig.~6 of \cite{cortines}). Another interesting study has been done
by Caruso {\it et al.} \cite{caruso} by using real earthquake data from the World and Northern California catalogs,
where they observed that the probability density of energy differences of subsequent earthquakes can also be well
fitted by a $q$-Gaussian with roughly the same value of $q$, i.e., $q\approx 1.75$ (see Fig.~2 of \cite{caruso}).
A more recent contribution along these lines consists in a molecular dynamical test of the $q$-CLT in a
long-range-interacting many-body classical Hamiltonian system known as HMF model \cite{HMFCLT}, where it was
numerically shown that, in the longstanding  quasi-stationary regime (where the system is only weakly chaotic),
the relevant densities appear to  converge to $q$-Gaussians with $q \approx 1.5$ (see for example Fig.~7
of \cite{HMFCLT}; see  also \cite{PluchinoRapisardaTsallis2008}). Moreover, $q$-Gaussians have also been
observed in the motion of Hydra cells in cellular aggregates \cite{UpadhyayaRieuGlazierSawada2001},
for defect turbulence \cite{DanielsBeckBodenschatz2004}, silo drainage of granular
matter \cite{ArevaloGarcimartinMaza2007}, cold atoms in dissipative optical lattices \cite{DouglasBergaminiRenzoni2006},
and dissipative 2D dusty plasma \cite{LiuGoree2008}. Finally, in a recent paper, we numerically investigated
the central limit behavior of deterministic dynamical systems \cite{tibet}, where one of our main purposes was
to see what kind of limit distributions emerge for the attractor whenever the dynamical system is not mixing
(for example at the edge of chaos, where the Lyapunov exponent vanishes) and thus the standard CLT is not
valid anymore. In \cite{tibet}, using the well-known standard example  of discrete one-dimensional dissipative
dynamical systems, the logistic map, defined as

\begin{equation} x_{t+1}=1-a x_t^2 \;\; ,
\label{logi}
\end{equation}
(where $0<a \le 2;\, |x_t| \le 1; \, t=0,1,2,...\,$), we numerically checked that, at the  edge of chaos
(i.e., close to the critical parameter value $a_c=1.401155189092...$), the tails of the limit distribution were
consistent with a $q$-Gaussian having, once again, a value of $q$ close to 1.75. However, the central part
of the distribution was not meticulously studied,
and neither was studied the precise dependence on the distance $a-a_c$ and on
the iteration number. In the present manuscript, our aim is to focus on these points,
having a closer look at sums of iterates of the logistic map close to its chaos threshold.

Although the iterates of a deterministic dynamical system can never be completely independent, one can still
prove some standard CLTs for such systems \cite{bill,beck,michael}, provided that the assumption of independent
identically distributed random variables is replaced by the  property that the system is sufficiently mixing
(i.e., asymptotic statistical  independence). As an example one can consider the logistic map at $a=2$ where it
is strongly mixing. For this system, it can be rigorously proved \cite{bill, beck} that the distribution of the
quantity

\begin{equation}
y:= \sum_{i=1}^N (x_i -\langle x \rangle)
\label{Y}
\end{equation}
becomes Gaussian for $N\to \infty$ after appropriate rescaling with a factor $1/\sqrt{N}$, regarding the initial
value $x_1$ as a random variable with a smooth probability distribution. Here $\langle x \rangle$ denotes the mean
of $x$, which happens to vanish for the special case $a=2$. This is a highly nontrivial result since the iterates
of the logistic map at $a=2$ are not independent but exhibit complicated higher-order correlations described by
forests of binary trees \cite{beck2}. Gaussian limit behavior is also numerically observed for other typical
parameter values in the chaotic region of the logistic map \cite{tibet}. Indeed, whenever the Lyapunov exponent
of the one-dimensional map is positive, one expects the CLT to be valid \cite{michael}.


Now we are ready to discuss the behavior of the logistic map at the edge of chaos for which a standard CLT is not
valid due to the lack of mixing. In order to calculate the average in Eq.~(\ref{Y}), it is necessary to
take the average over a large  number of $N$ iterates as well as a large number $n_{ini}$ of randomly chosen initial
values $x_1^{(j)}$, namely,

\begin{equation}
\langle x \rangle = \frac{1}{n_{ini}} \frac{1}{N} \sum_{j=1}^{n_{ini}} \sum_{i=1}^N x_i^{(j)}\; .
\end{equation}
These conditions are important due to potentially non-ergodic and non-mixing
behavior.

In principle, at the edge of chaos, taking $N\to \infty$ is not
the only ingredient for the system to attain its limit
distribution. It is also necessary to localize the critical point
(the chaos threshold) with infinite precision. In other words,
theoretically, for a full description of the shape of the
distribution function on the attractor, one needs to take the
$a_c$ value with infinite precision  as well as taking $N\to
\infty$. On the other hand, in numerical experiments, neither the
precision of $a_c$ nor the $N$ values can approach infinity. In
fact, numerically one can see the situation as a kind of
interplay between the precision of $a_c$ and the number $N$ of
iterates. For a given finite precision of $a_c$ (slightly above
the exact critical value), if we use a very large $N$, then the
system quickly feels  that it is not exactly at the chaos
threshold, and the central part of its probability distribution
function typically becomes a Gaussian (with only small deviations in the
tails). On the other hand, for the same distance to $a_c$, if we take $N$ too
small, then the summation given by Eq.~(\ref{Y}) starts to be
inadequate to approach the edge-of-chaos limiting distribution,
and the central part of the distribution around zero exhibits peaks. This is indeed a direct consequence of the
fact that the attractor of the system at the edge of chaos is a
fractal that only occupies a tiny part of the full phase space
(see \cite{robledo} for details). This is the reason why the
central parts of the distributions shown in \cite{tibet} do not
present the typical smooth shape of $q$-Gaussians. Indeed, the
values of $N$ chosen in \cite{tibet} ($2^{14}$ and $2^{15}$) are
too small for the precision of $a_c$ (1.401155189092) to obtain a
complete picture of the entire  distribution including both
central parts and tails. On the other hand, for the above
precision of $a_c$, one can think about numerical experiments
with $N$ values at the level of, say, $2^{40}$ or more, for which
the central part would approach a Gaussian since the system
starts to realize that it is not exactly at the edge of chaos. We
observe that between these two extremes, for a given precision of
$a_c$, there exists a range of values of $N$ for which the
probability density of the system is well approximated by a
$q$-Gaussian in the entire region. Unfortunately, these kinds of
large $N$ values which are necessary to fully verify this
observation if we approach $a_c$ with say 12 digits precision (as we
did in \cite{tibet}) cannot normally be reached in numerical
experiments. However, we can check this scenario using less
precision for $a_c$. This will in turn make the appropriate $N$
value become small enough so that we are able to handle the
numerics with standard computers.

\begin{figure}
\epsfig{file=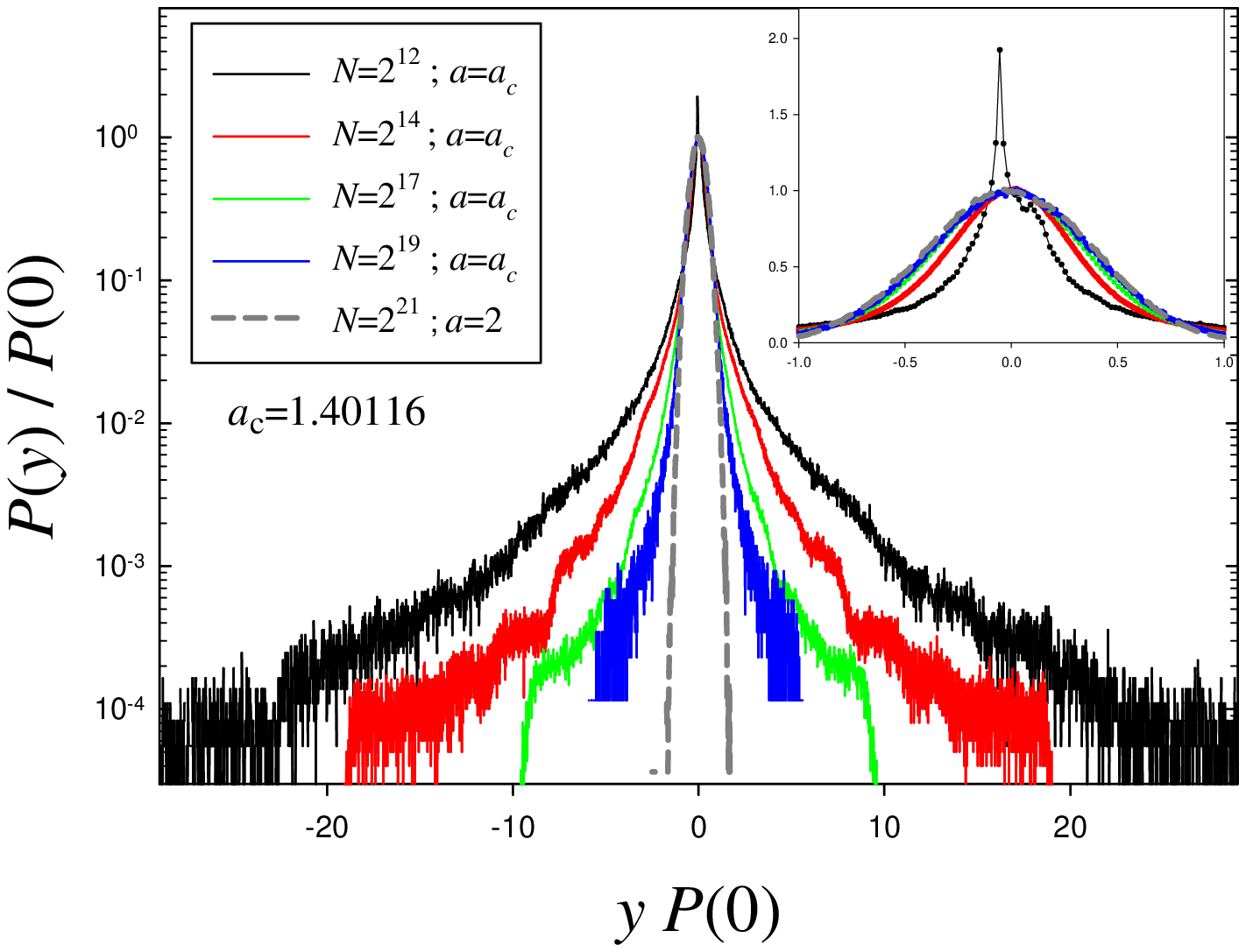} 
\caption{Probability density function
$P(y)$ of the quantity $y$ rescaled by $P(0)$. The map is close to the
edge of chaos with $5$ digits  precision ($a=1.40116$). The
tendency to approach the Gaussian is evident as $N$ increases and
$a$ is kept fixed.}
\end{figure}

As a representative illustration, we first focus on an $a$ value
in the vicinity of the critical point with 5 digits precision
($a=1.40116$). For this case (and several other cases which are
not described here), we numerically verified the above-mentioned
scenario, as can be seen in Fig.~1. In our simulations, after
omitting a transient of the first $2^{12}$ iterates 
(we checked that the results are independent of the omitted transient length  
as long as it is large enough), we calculated
the quantity $y$ in Eq.~(\ref{Y}) for various $N$ values and
obtained its probability distribution from an ensemble of uniformly
distributed initial values. It is worth mentioning that a similar
picture emerges for almost any vicinity of the critical point, but of
course with different $N$ values. It is clearly seen from the
figure that, for $a=1.40116$, $N=2^{19}$ is so large
that the density approaches a Gaussian in the central region with
small deviations in the tails (further increase of $N$ values
would make the whole curve become a Gaussian), whereas for
$N=2^{12}$ the curve has heavy tails and a peaked central part
(this curve can be fitted neither by a Gaussian nor by a
$q$-Gaussian in the entire region). On the other hand, between
these two extreme cases, there is an appropriate range of $N$
values (around $2^{17}$ for this example of $a-a_c$)
for which the distribution is consistent with a $q$-Gaussian of the
form

\begin{equation}
P(y) \sim e_q^{-\beta y^2} := \frac{1}{(1+\beta (q-1) y^2)^{\frac{1}{q-1}}} \;,
\label{qexp}
\end{equation}
(where $q$ and $\beta$ are suitable parameters) for the entire region.

Let us now provide a theoretical argument what the optimum value of $N$ could be to achieve best convergence
to a $q$-Gaussian. Finite precision of $a_c$ means that the parameter $a$ of the system is at some distance
$|a-a_c|$ from the exact critical point $a_c=1.401155189092...$. Suppose we are slightly above the critical
point ($a>a_c$), by an amount

\begin{equation}
|a-a_c| \sim \frac{1}{\delta^n},
\label{distance}
\end{equation}
where $\delta = 4.6692011...$ is the Feigenbaum constant. Then there exist $2^n$ chaotic bands of the attractor
with a selfsimilar structure, which approach the Feigenbaum attractor for $n \to \infty$ by the band splitting
procedure (see e.g. \cite{beck-schloegl}, p.10, for more details). Suppose we perform $2^n$ iterations of the map
for a given initial value with a parameter $a$ as given by Eq.~(\ref{distance}). Then after $2^n$ iterations we
are basically back to the starting value, because we fall into the same band of the band splitting structure.
This means the sum of the iterates $\sum_{i=1}^{2^n} x_i$ will essentially approach a fixed
value $w =2^n \langle x \rangle$ plus a small correction $\Delta w_1$ which describes the small fluctuations of
the position of the $2^n$th iterate within the chaotic band. Hence

\begin{equation}
y_1=\sum_{i=1}^{2^n} (x_i-\langle x \rangle ) =\Delta w_1.
\end{equation}
If we continue to iterate for another $2^n$ times, we obtain

\begin{equation}
y_2=\sum_{i=2^n+1}^{2^{n+1}} (x_i-\langle x \rangle )= \Delta w_2.
\end{equation}
The new fluctuation $\Delta w_2$ is not expected to be
independent from the old one $\Delta w_1$, since correlations of
iterates decay very slowly if we are close to the critical point.
Continuing, we finally obtain

\begin{equation}
y_{2^n}= \sum_{i=4^n-2^n+1}^{4^n} (x_i- \langle x \rangle )=\Delta w_{2^n}
\end{equation}
if we iterate the map $4^n$ times in total.  The total sum of iterates

\begin{equation}
y=\sum_{i=1}^{4^n} (x_i -\langle x \rangle )=\sum_{j=1}^{2^n} \Delta w_j
\end{equation}
can thus be regarded as a sum of $2^n$ strongly correlated random variables $\Delta w_j$, each being influenced by
the structure of the $2^n$  chaotic bands at distance $a-a_c\sim \delta^{-n}$ from the Feigenbaum attractor.
There is a 1-1 correspondence between these $2^n$ random variables $\Delta w_j, j=1,\ldots ,2^n$ and the $2^n$ chaotic
bands of the attractor, which remains preserved if $n$ is further increased. It is now most reasonable to assume
that the above system of a sum of $2^n$ correlated random variables $\Delta w_j$ exhibits data collapse (and hence
convergence to a well-defined limit distribution) under successive renormalization transformations
$n \to n+1 \to n+2 \to n+3 \cdots$. The limit distribution may indeed be a $q$-Gaussian, as indicated by our
numerical experiments. The above scaling argument implies that the optimum iteration time $N^*$ to observe convergence
to a $q$-Gaussian limit distribution is given by

\begin{equation}
N^*\sim 2^{2n} \label{NNN}
\end{equation}
where, at a given distance $a-a_c$, the number $n$ is given by

\begin{equation}
n \approx - \frac{\log |a-a_c|}{\log \delta}.
\label{nnn}
\end{equation}

Another way to formulate our scaling argument is as follows.
Consider a given distance from the critical point where there are
$2^n$ chaotic bands. The relevant function in the Feigenbaum
renormalization scheme is the $2^n$-th iterated function
$f_*:=f^{2^n}$ rather than the original function $f$ itself. The
iterates of $f_*$ will be highly correlated. Let $k$ be the number
of iterates of $f_*$ that are being added up. For $k>>2^n$ we get
a Gaussian distribution for the probability distribution of the
sum, since the system feels that it is in the chaotic regime
$a>a_c$. For $k<<2^n$ there is no chance to get anything smooth
for the probability distribution of the sum since the iteration
number is too small. The interesting intermediate case is the
case $k=2^n$. Here, due to the strong correlations, there is the
chance to get a smooth $q$-Gaussian limit distribution. Since in
this case the number of chaotic bands is the same as the number
of iterates of $f_*$ that are being added up, the theory is
invariant under successive renormalization transformations $n\to
n+1 \to n+2 ...$. But iteration number $k=2^n$ for $f_*$
corresponds to iteration number $N^*=2^{2n}$ for the original map,
which is our equation~(\ref{NNN}).

\begin{table}
\caption{\label{tab:Table}Various $a$ values used in the
simulations and the associated values of $n$ and $N^*$.}
\begin{ruledtabular}
\begin{tabular}{||c|c|c|c||} \hline
  $a$        &  $|a-a_c|$           &  $n$    &  $N^{*}$    \\ \hline
$1.40209$    & $9.348\cdot 10^{-4}$ & $4.526 {\simeq 9/2}$ &  $2^9$    \\ \hline
$1.401588$   & $4.328\cdot 10^{-4}$ & $5.026{\simeq 10/2}$ &  $2^{10}$ \\ \hline
$1.401354$   & $1.988\cdot 10^{-4}$ & $5.531{\simeq11/2}$ &  $2^{11}$ \\ \hline
$1.401248$   & $9.281\cdot 10^{-5}$ & $6.025{\simeq 12/2}$ &  $2^{12}$ \\ \hline
$1.401198$   & $4.281\cdot 10^{-5}$ & $6.527{\simeq 13/2}$ &  $2^{13}$ \\ \hline
$1.401175$   & $1.981\cdot 10^{-5}$ & $7.027{\simeq 14/2}$ &  $2^{14}$ \\ \hline
$1.4011644$  & $9.211\cdot 10^{-6}$ & $7.524{\simeq 15/2}$ &  $2^{15}$ \\ \hline
$1.40115945$ & $4.261\cdot 10^{-6}$ & $8.025{\simeq 16/2}$ &  $2^{16}$ \\ \hline
$1.40115716$ & $1.971\cdot 10^{-6}$ & $8.525{\simeq 17/2}$ &  $2^{17}$ \\ \hline
\end{tabular}
\end{ruledtabular}
\end{table}

In order to check these types of scaling arguments, we numerically
studied various $a$ values as listed in the Table. 
{First we show in Fig.~2a and 2b the densities obtained for the case $N=N^*=2^{2n}$ for 
each value of $a$, where $2n$ is odd and even, respectively.
Three important aspects are evident: \\
(i)~the curves obtained for these cases exhibit a very clear data collapse; \\
(ii)~the envelopes of the histogram data can be well fitted {\it everywhere} (i.e. both in the tails and in the
central region) by a $q$-Gaussian given by Eq.(\ref{qexp}) with
$q=1.68$ if $2n$ is odd and $q=1.70$ if $2n$ is even, for the respective $a$ 
values given in the Table. It is also evident from this figure that small log-periodic 
modulations are present on top of the curves. The existence
of such modulations is expected due to discrete 
scale invariance in these types of systems, as demonstrated long ago by de 
Moura et al. \cite{moura} (see also \cite{RobledoMoyano2008});\\
(iii)~as we approach $a_c$ more closely, and consequently the appropriate
value $N^*$ increases, the region consistent with the $q$-Gaussian
grows in size. In order to illustrate this statement we include Fig.~3
which shows the same data in a different representation. 
One can better see there how the $q$-Gaussians 
with log-periodic-like modulations develop as $a$ approaches $a_c$.
$q$-Gaussians correspond to a straight line in these types of plots. 
We also determined and show in Fig.~4 how $P(0)$ evolves with $N$ for the cases studied in Fig.~2.
}

\begin{figure}
\includegraphics*[height=9cm]{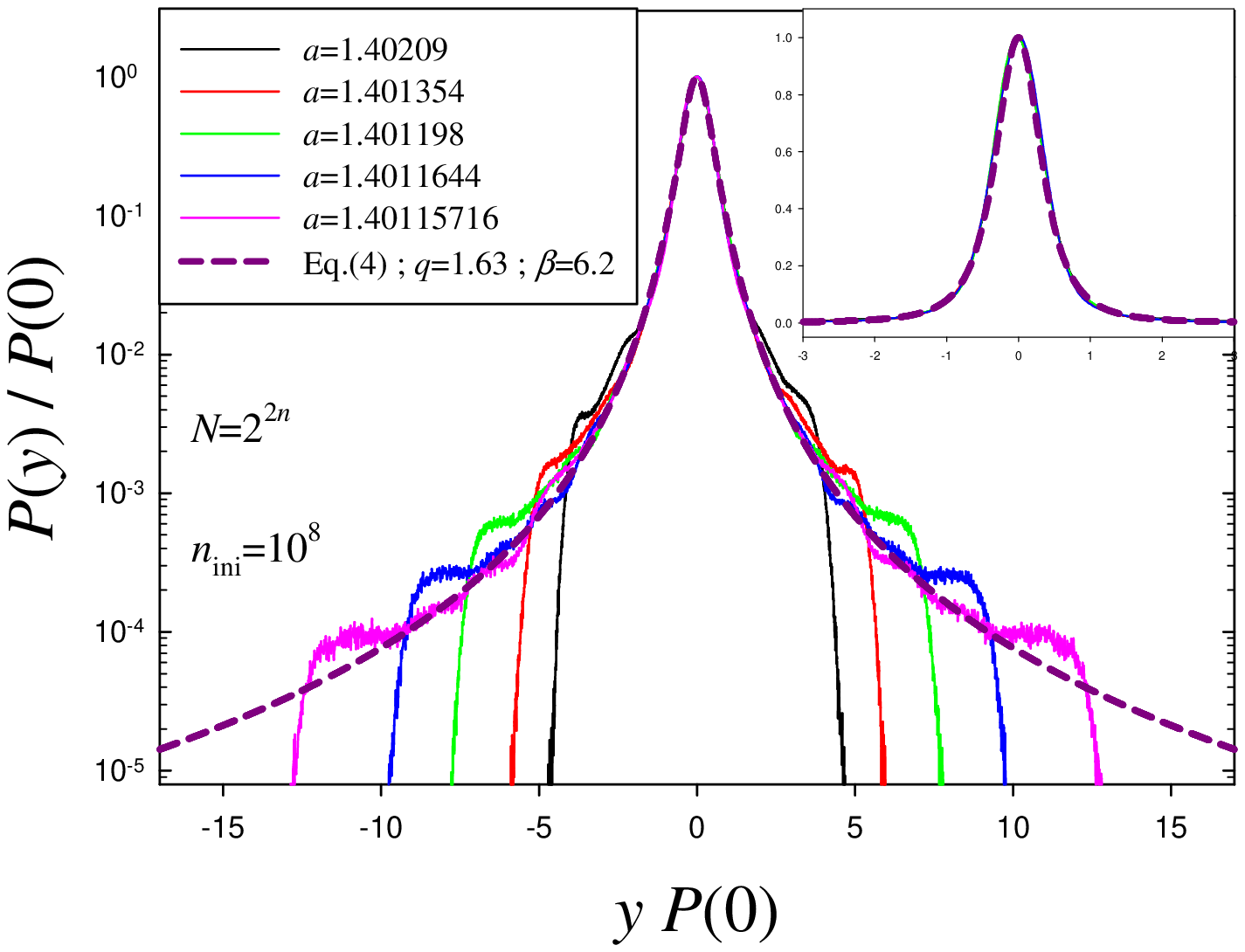}
\includegraphics*[height=9cm]{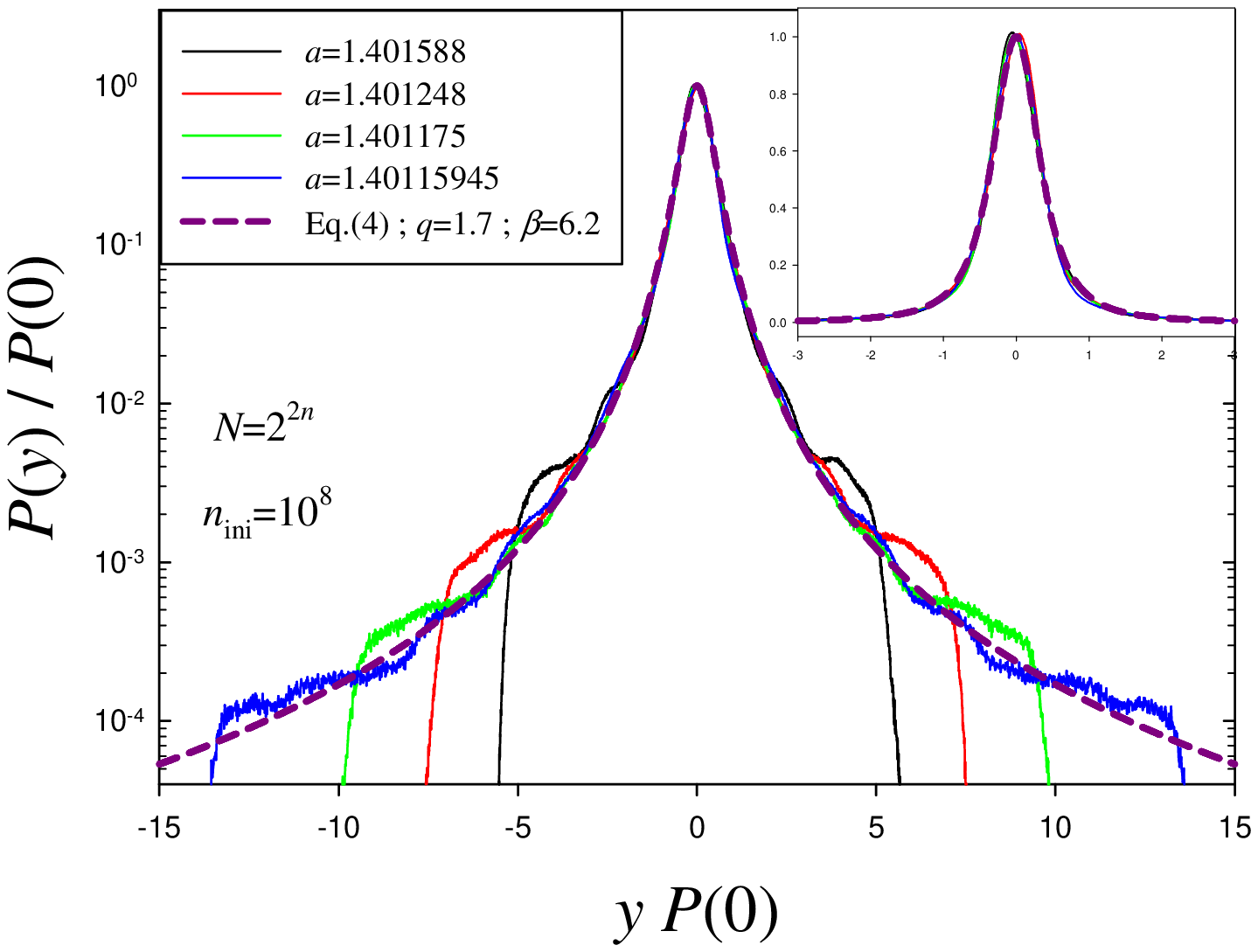}
\caption{\label{fig:Fig2} Data collapse of probability density
functions for the cases $N=2^{2n}$, where $2n$ is odd (a) and
even (b). As $n$ increases, a good fit using a $q$-Gaussian with
$q=1.68$ and $\beta=6.2$ (a) and $q=1.70$ and $\beta=6.2$ (b) 
is obtained for regions of increasing size. 
Inset: The linear-linear plot of the data for a better visualization 
of the central part.}
\end{figure}

\begin{figure}
\includegraphics*[height=9cm]{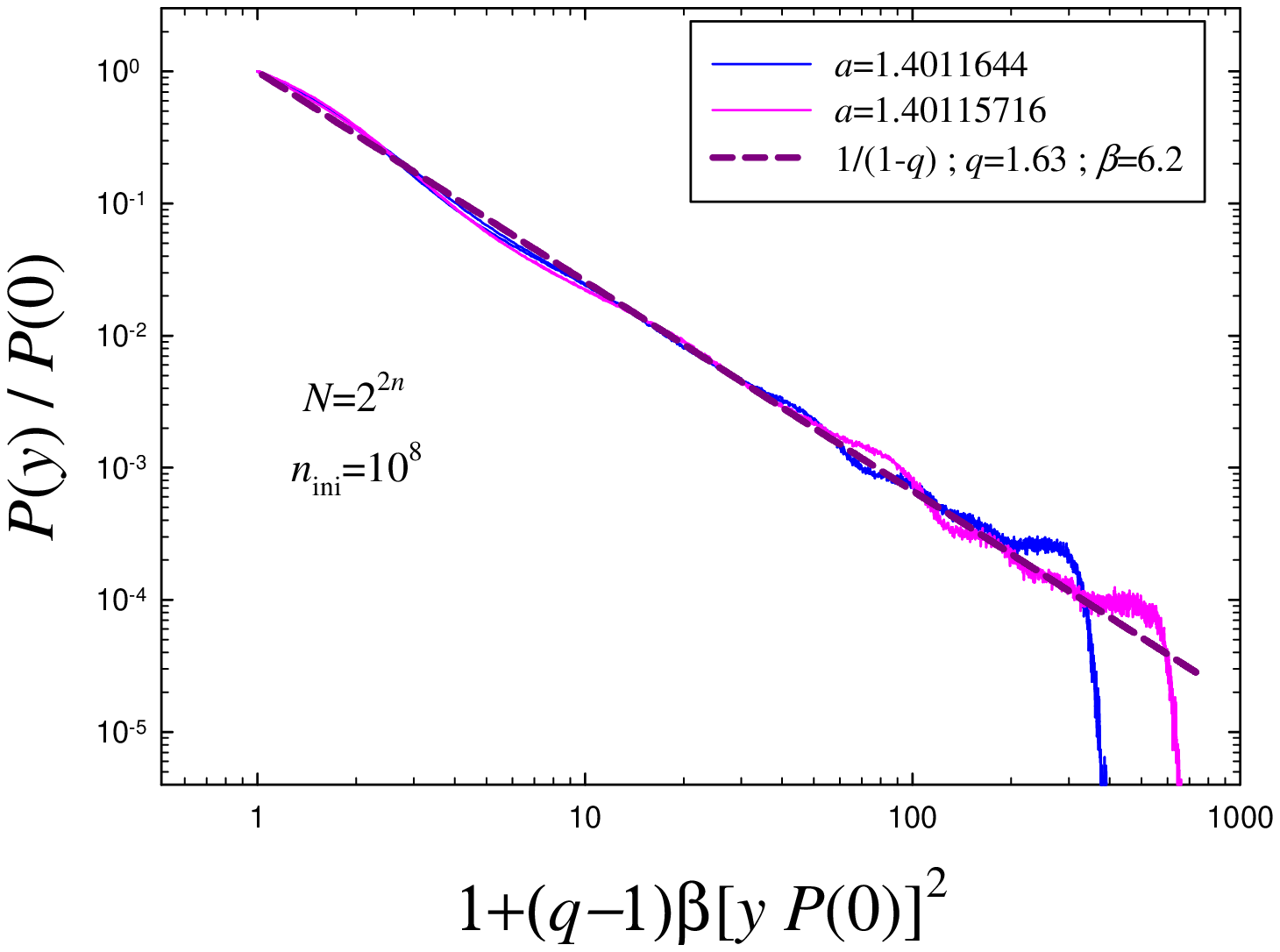}
\includegraphics*[height=9cm]{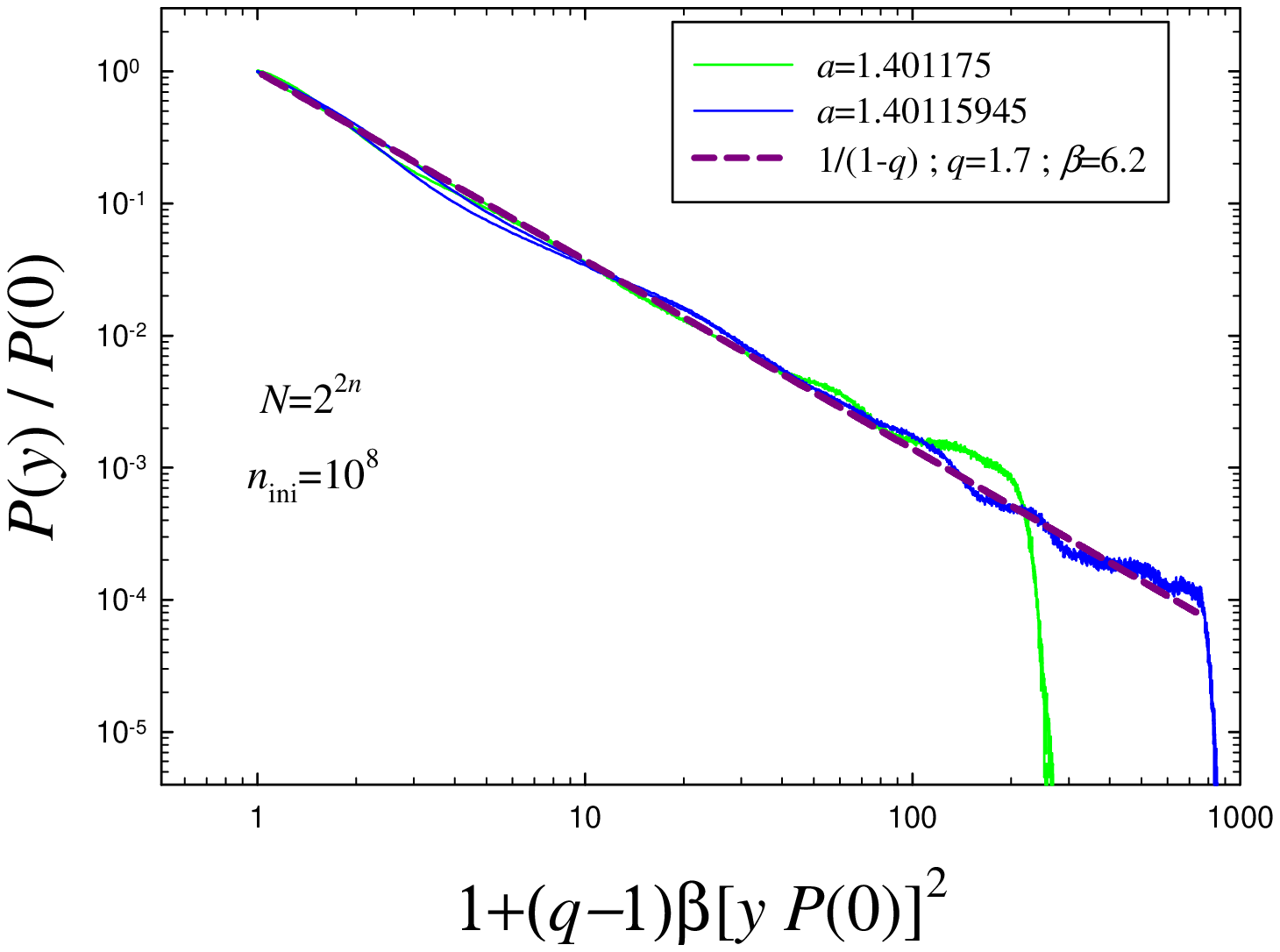}
\caption{\label{fig:Fig3} Probability density functions plotted against 
$1+(q-1)\beta [y P(0)]^2$ on a log-log plot for the cases $N=2^{2n}$, 
where $2n$ is odd (a) and even (b). 
A straight line is expected with a slope $1/(1-q)$ if the curve is a 
$q$-Gaussian. It is clearly seen how the straight line is surrounded 
by the log-periodically modulated curves.}
\end{figure}

\begin{figure}
\epsfig{file=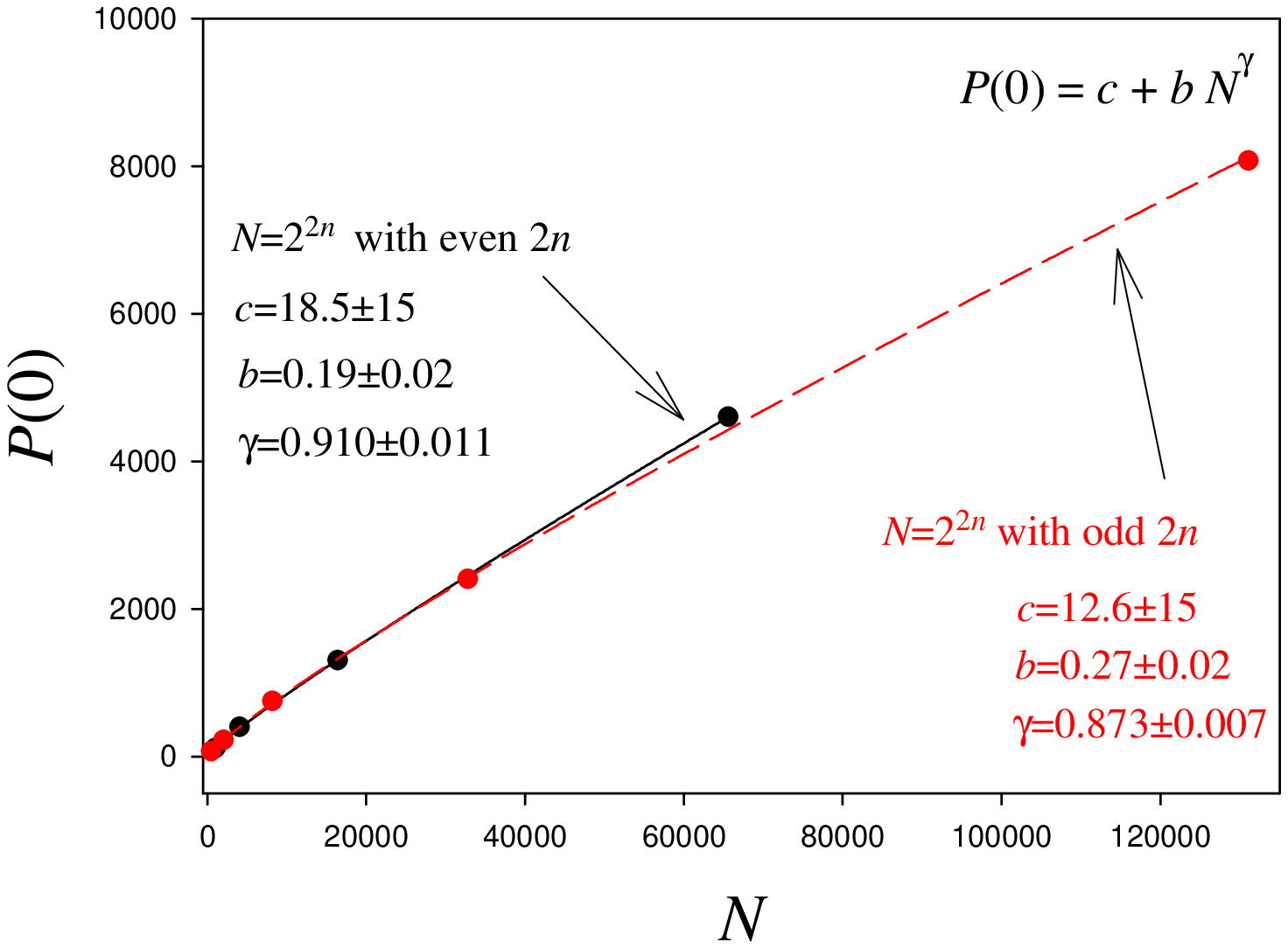} 
\caption{\label{fig:Fig4} $N$-dependence of
the rescaling factor $P(0)$ that yields data collapse as displayed
in Fig.~2a and 2b.}
\end{figure}

In Fig.~5 we plot the $q$-logarithm (defined to be the inverse
function of the  $q$-exponential given in Eq.~(\ref{qexp}),
namely $ln_q(x)=(x^{1-q}-1)/(1-q)$) of the  same data as in
Fig.~2a. This provides a further visualization of the aspect
already mentioned in (iii), namely that as the precision of $a_c$
and the value of $N$ increase, the region consistent with a
$q$-Gaussian extends in size. There is a clear numerical
indication that $q$-Gaussians are a good approximation of the data 
if both the precision of $a_c$ and the value of $N$
go to infinity.

\begin{figure}
\epsfig{file=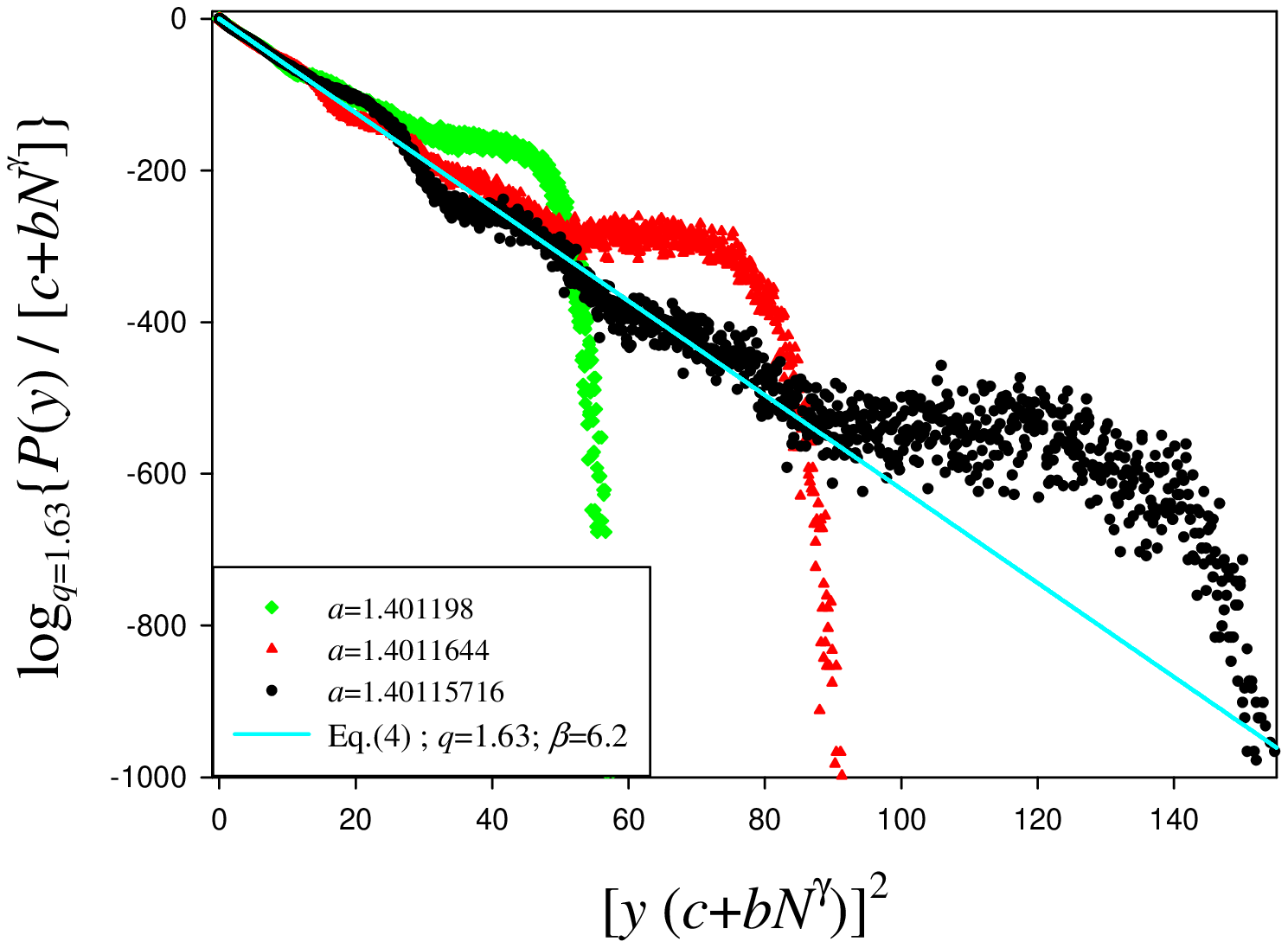}
\caption{\label{fig:Fig5}
$q$-logaritmic plot of the same data as in Fig.~2a. As the distance to $a_c$
descreases
and the value of $N$ increases, it is clearly seen that the region consistent with
(modulated) $q$-Gaussian behavior is widening.}
\end{figure}

{We also checked that our results are not induced by roundoff or
similar numerical artifacts. Throughout our simulations we used 
double precision of Intel Fortran, achieving good statistics 
($n_{ini}=10^8$ for all cases). We also tested our results
with less statistics but using higher precision 
(quadruple precision of Intel Fortran).
Since no significant differences were observed,
we used double precision with high statistics in most of our simulations, which 
allowed us to see the observed $q$-Gaussians to be modulated by 
small log-periodic-like oscillations. 
The detailed structure of these oscillations depends on the precise value of $a$ in a very complicated way,
which is to be expected, since the logistic map is well-known
to exhibit very complex behavior as a function of $a$.
Nevertheless, our numerical experiments provide evidence that the envelope of the data 
is always very well approximated by a q-Gaussian, provided the typical number of 
iterations is chosen according to the scaling relations given by Eqs. (\ref{NNN}) and (\ref{nnn}). 
Another interesting subject is, no doubt, to analyse the cases $N<<N^*$ and $N>>N^*$ 
in order to understand the crossover phenomena from the peaked region to $q$-Gaussian region 
and finally to the normal Gaussian region. This will be addressed elsewhere in the 
near future.
}


Summarizing, we have presented numerical evidence that the distributions of sums of iterates 
of the logistic map in a close vicinity of the edge of chaos are well approximated by $q$-Gaussian 
probability distributions, provided the typical number of iterations scale in line with 
Eqs.~(\ref{NNN}) and (\ref{nnn}). This illustrates the strongly {\it correlated} nature of this 
paradigmatic nonlinear dynamical system (which models a great variety of more complex physical situations, 
as it is well-known in the literature). The $q$-Gaussians are precisely the limit distributions of 
an important class of strongly correlated random systems in the realm of the recently proved
$q$-generalized Central Limit Theorem. This feature, together with the fact that they optimize 
within appropriate constraints the nonadditive entropy $S_q$, is of interest for the mathematical 
foundations of nonextensive statistical mechanics, as well as for many
real-world problems that consist of sums of correlated random variables.
Needless to say that the analytical proof of the results numerically
obtained here would be most welcome. 
As open questions we may mention (i) the careful study of other dissipative
and conservative maps, starting with the one-dimensional $z$-logistic one, 
and (ii) the investigation of the precise convergence radius to the $q$-Gaussian 
limit form and its oscillating small correction
terms as a function of $N$ and $a-a_c$. A full numerical study of these
points certainly requires high computational power.


We would like to thank H.J. Hilhorst for very fruitful discussions.
This work has been supported by TUBITAK (Turkish Agency) under the Research Project number 104T148.
C.T. acknowledges partial financial support from CNPq and Faperj (Brazilian Agencies).

\end{document}